\documentclass{aa}

\usepackage{natbib}
\usepackage{graphicx}

\usepackage{epsfig}
\usepackage{amssymb}
\usepackage{amsmath}
\usepackage{txfonts}
\bibpunct{(}{)}{,}{}{}{,}

\begin{document}

\title{Long-term trends of magnetic bright points:}
\subtitle{I. Number of MBPs at disc centre}
\titlerunning{Number of MBPs at disc centre}

\author{D. Utz
\inst {1, 2}
\and R. Muller
\inst 3
\and S. Thonhofer
\inst {1,2}
\and A. Veronig
\inst 1
\and A. Hanslmeier
\inst {1, 3}
\and M. Bodn\'arov\'a
\inst {4}
\and M. B\'{a}rta
\inst {5}
\and J. C. del Toro Iniesta
\inst {2}
}

\institute{IGAM/Institute of Physics, University of Graz, Universit{\"a}tsplatz 5, 8010 Graz, Austria
\and Instituto de Astrof\'{i}sica de Andaluc\'{i}a (CSIC), Apdo. de Correos 3004, 18080 Granada, Spain
\and Laboratoire d'Astrophysique de Toulouse et Tarbes, UMR5572, CNRS et Universit{\'e} Paul Sabatier Toulouse 3, 57 avenue d'Azereix, 65000 Tarbes, France
\and Astronomical Insitute, Slovak Academy of Sciences, 05960 Tatranska Lomnica, Slovakia
\and Astronomical Institute of the Czech Academy of Sciences,   Fri\v{c}ova 298,25165 Ond\v{r}ejov, Czech Republic}

\date{Received 19 February 2015 /
Accepted 1 October 2015 }
\abstract{The Sun shows an activity cycle that is caused by its varying global magnetic field. During a solar cycle, sunspots, i.e. extended regions of strong magnetic fields, occur in activity belts that are slowly migrating from middle to lower latitudes, finally arriving close to the equator during the cycle maximum phase. While this and other facts about the strong extended magnetic fields have been well known for centuries, much less is known about the solar cycle evolution of small-scale magnetic fields. Thus the question arises if similar principles exist for small-scale magnetic fields.} {To address this question, we study magnetic bright points (MBPs) as proxies for such small-scale, kG solar magnetic fields. This study is based on a homogeneous data set that covers a period of eight years. The number of detected MBPs versus time is analysed to find out if there is an activity cycle for these magnetic features too and, if so, how it is related to the sunspot cycle.
} {An automated MBP identification algorithm was applied to the synoptic Hinode/SOT G-band data over the period November 2006 to August 2014, i.e. covering the decreasing phase of Cycle 23 and the rise, maximum, and early decrease of Cycle 24. This data set includes, at the moment of investigation, a total of 4 162 images, with about 2.9 million single MBP detections. 
} {After a careful preselection and monthly median filtering of the data, the investigation revealed that the number of MBPs close to the equator is coupled to the global solar cycle but shifted in time by about 2.5 years. Furthermore, the instantaneous number of detected MBPs depends on the hemisphere, with one hemisphere being more prominent, i.e. showing a higher number of MBPs. After the end of Cycle 23 and at the starting point of Cycle 24, the more active hemisphere changed from south to north. Clear peaks in the detected number of MBPs are found at latitudes of about $\pm 7^{\circ}$, in congruence with the positions of the sunspot belts at the end of the solar cycle.} {These findings suggest that there is indeed a coupling between the activity of MBPs close to the equator with the global magnetic field. The results also indicate that a significant fraction of the magnetic flux that is visible as MBPs close to the equator originates from the sunspot activity belts. However, even during the minimum of MBP activity, a percentage as large as 60\% of the maximum number of detected MBPs has been observed, which may be related to solar surface dynamo action.}


\maketitle

\section{Introduction}
Since the pioneering work of \citet{1844AN.....21..233S}, it is well known that the Sun has a sunspot or activity cycle and is thus a variable star. \citet[][]{1908ApJ....28..315H} was the first to establish that sunspots are actually created by strong and extended magnetic fields. Thus the varying solar activity is closely correlated to the Sun's global magnetic field, which shows a roughly 22-year magnetic cycle, comprised of two 11-year sunspot cycles \citep[e.g.][]{1961ApJ...133..572B}. Within such a cycle, sunspots are formed in so-called active latitude belts, which over the course of a solar cycle migrate closer and closer to the equator \citep[see][]{1858MNRAS..19....1C,2003ApJ...589..665H}. Furthermore sunspots obey several other rules, such as Joy's law of the tilting of sunspot groups and Hale's magnetic polarity rules \citep[the details of which can be found in the original paper of][]{1919ApJ....49..153H}.  
It is known that the magnetic field starts to be torn out from the sunspots and is then transported via the meridional large-scale flows to the polar regions of the Sun \citep[e.g.][]{1961ApJ...133..572B,1966ApJ...144..723S,1995A&A...303L..29C} where, with some temporal lag,  it causes  the polar magnetic field reversals \citep[for recent progress in observations of the polar magnetic fields see, e.g.][]{2008ApJ...688.1374T,2012ApJ...753..157S}. This transported magnetic flux consists of small-scale features whose properties and behaviour along a whole solar cycle was of interest already in the 80's of the last century \citep[e.g.][]{1982SoPh...80...15L}. Those earlier investigations yielded different and partly contradicting results. Some studies found a correlation between the magnetic fields and the sunspot cycle, while others  report that small-scale magnetic fields seemed to be anti-correlated with the sunspot cycle. Finally, there have been studies that state  there is no correlation at all. For a discussion of these results, we refer the reader to \citet[][]{2011ApJ...731...37J} and references therein. For more detail on the topic of the solar cycle we refer to the review by \citet{2010LRSP....7....1H}.

All of these observational findings have to be explained by a dynamo theory \citep[see the review of][and references therein]{2010LRSP....7....3C}. In  recent years, numerical simulations of the dynamo process \citep[e.g.][]{2015arXiv150506632K}, as well as of the evolution of the magnetic network during a solar cycle \citep[e.g.][]{2014ApJ...796...19T},    agree with these observational findings to a more and more satisfying
degree but there are still major open questions to be addressed \citep[e.g. the final chapter of][]{2010LRSP....7....3C}.  
\begin{figure*}
\centering
\includegraphics[width=0.92\textwidth]{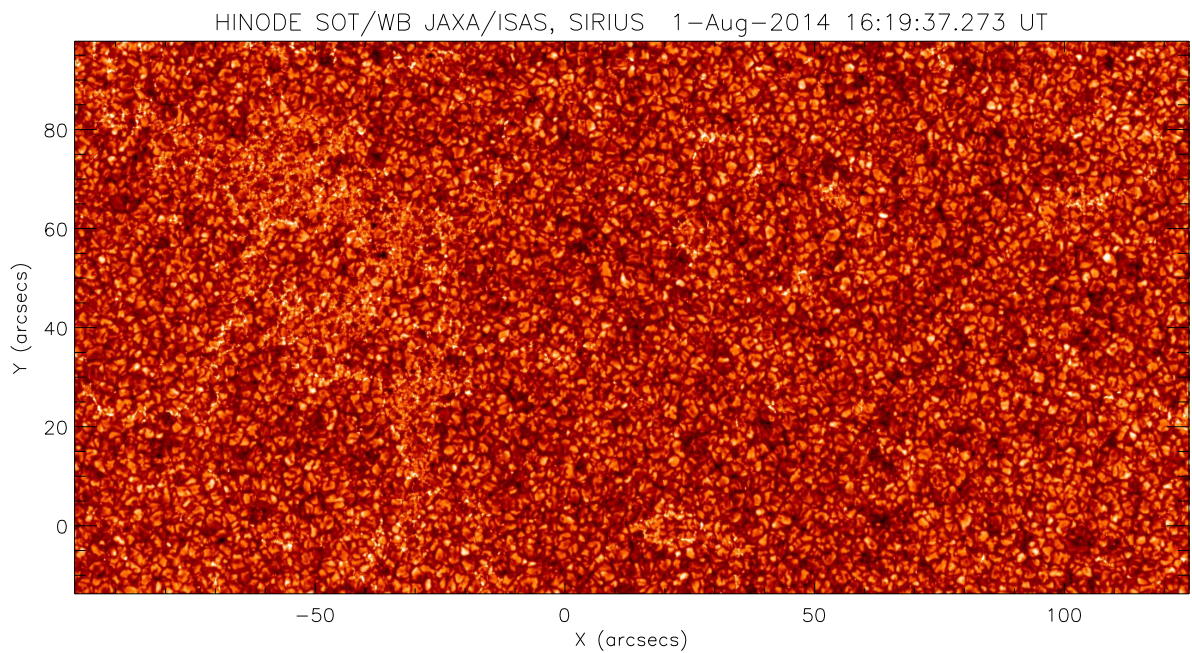}
\caption{Synoptic Hinode/SOT G-band filtergram from 1 August 2014. To increase the contrast of the image, the data were clipped to be between 33\% and 75\% of the image's intensity maximum.
}
\label{figure1}
\end{figure*}

These theories and models take in most cases only the strong and extended magnetic fields into account. Thus the following questions arise: what happens to the magnetic network \citep[see, for example, previous studies performed with SOHO/MDI by][]{2011ApJ...731...37J,2012ApJ...745...39J} and   to proxies of small-scale kG concentrations that are often present within this network and known as network bright points \citep{1983SoPh...85..113M} or  magnetic bright points \citep[MBPs; see, e.g.][and, for more general reviews on small-scale magnetic fields, \citealt{1993SSRv...63....1S} or \citealt{2009SSRv..144..275D}]{2004A&A...428..613B}. MBPs are representations of strong magnetic field concentrations \citep[see, e.g.][]{2001ApJ...553..449B,2007A&A...472..607B,2013A&A...554A..65U} and often reach more than 1~kG\footnote{An ongoing debate actually exists on whether all MBPs  have to have magnetic field strengths above 1~kG or not, with theoretical indications that they should have more than 1~kG \citep[see][]{2014A&A...562L...1C,Riethmueller2014}.}. They feature small diameters in the range of a few hundred km \citep[e.g.][]{2004A&A...422L..63W,2014RAA....14..741Y}. The lower limit of the size distribution has most likely not yet been discovered  as with increasing telescope apertures the measured lower limit of the size distribution of the MBPs is still decreasing \citep[e.g.][]{2010ApJ...725L.101A}. Finally, MBPs are characterised by their increased intensity, compared to the dark intergranular lanes in which they are situated \citep[e.g.,][]{1983SoPh...87..243M,2004A&A...422L..63W,2007ApJ...661.1272B}. This is especially true for molecular bands like the G band. Therefore this spectral region is often chosen when identifying them \citep{1984SoPh...94...33M,2001A&A...372L..13S,2003ApJ...597L.173S}. Recent studies, based on observations and computer simulations, have investigated the temporal evolution from the formation until the disintegration of single features, which occur over timescales of just a few minutes \citep{2013SoPh..tmp..276B,2014A&A...565A..84H,2014ApJ...796...79U}. 

\begin{figure}
\begin{center}
\includegraphics[width=0.95\columnwidth]{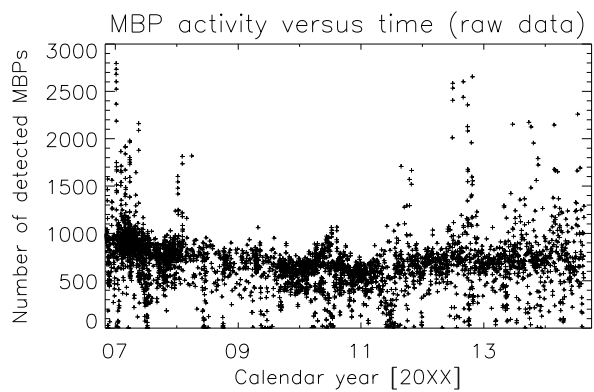}
\end{center}
\caption{Detected number of MBPs versus time using all available G-band images of the Hinode synoptic programme.}
\label{figure2}
\end{figure}

However, almost no long-term studies exist regarding MBPs \citep[a first attempt was undertaken by][but with a much smaller image sample per year and less homogeneous data compared to this study]{1984SoPh...94...33M}. Among the few studies which try to shed some light on this topic are \citet[][]{2011SoPh..274...87M} and \citet[][]{2015Hvar}, stating some preliminary results with regard to the current investigation. This is caused by the lack of suitable and stable long-term data sets that cover the solar cycle related variations of MBPs. Thus the following questions remain: 1) Are there laws governing the occurrence frequency and characteristics of small-scale magnetic fields that are similar to the established rules for sunspots \citep[see, e.g][]{2003A&A...405.1107M,2011ApJ...731...37J}? 2) Are these small-scale magnetic fields independent of the global magnetic field cycle \citep[e.g.][]{1982SoPh...80...15L,2013A&A...555A..33B}? 3) Do completely different rules apply for small-scale magnetic fields that  might be dependent on the strength of the cycle \citep[e.g.][]{2015arXiv150506632K}? Answering these questions will not only improve our understanding of the solar cycle but also that of the underlying dynamo theories, as there is still a debate about the possibility of an acting surface dynamo being responsible for the formation of small-scale solar magnetic fields \citep[see][]{1993A&A...274..543P,2012ASPC..455....3L,2014ApJ...789..132R}. 

New possibilities for answering these questions on long-term evolutions have  opened up in  recent years owing to space missions, since observations from instruments flown in space are not subject to varying atmospheric conditions and thus their behaviour is, in general, more stable \citep[see, among others,][]{2003ApJ...584.1107H,2011ApJ...731...37J,2014SoPh..289.3371B,2014PASJ...66S...4L}. For the purposes of long-term investigations on small-scale solar magnetic fields, the mission of choice is Hinode. Operating now for more than eight years, it delivers stable, seeing-free, and homogeneous data with high spatial resolution that can provide insights into the long-term behaviour of small-scale magnetic fields.

\begin{figure}
\begin{center}
\includegraphics[width=0.95\columnwidth]{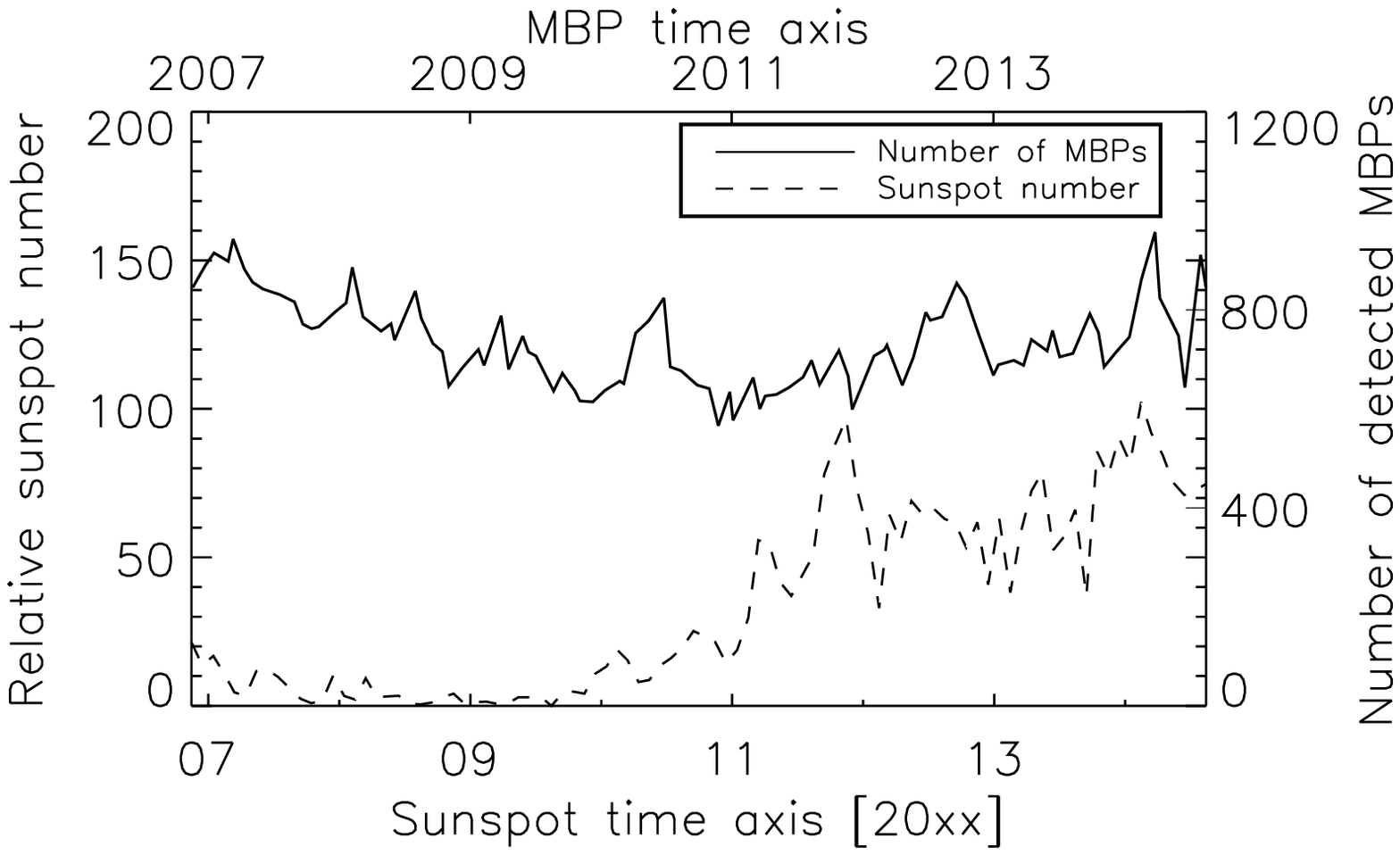}
\includegraphics[width=0.95\columnwidth]{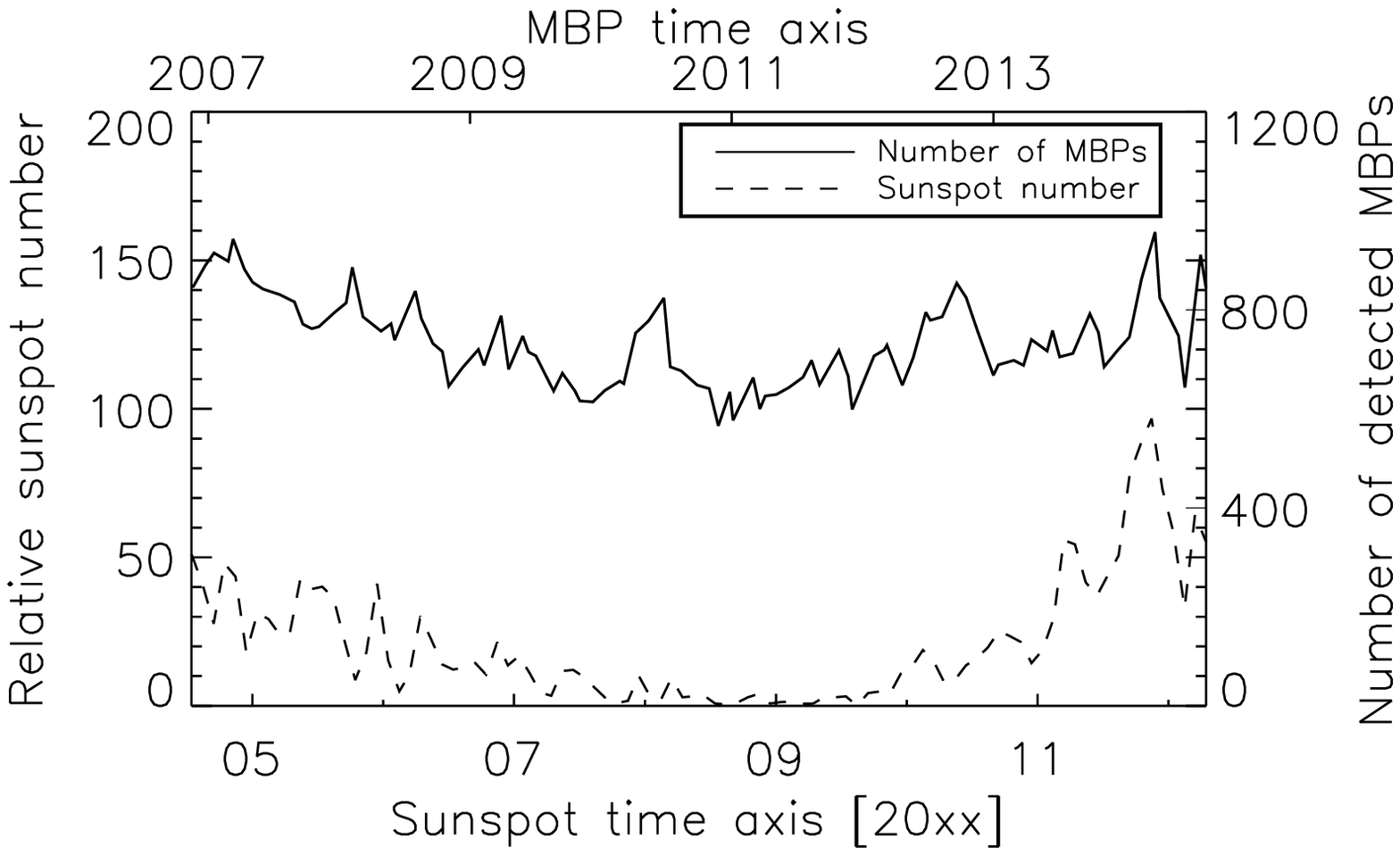}
\end{center}
\caption{Top panel: Monthly median number of detected MBPs after image selection via the contrast criterion versus time (solid line). The dashed line shows the relative sunspot number obtained from the Sunspot Index and Long-term Solar Observations Center. Bottom panel: As the top panel, but for the relative sunspot number shifted by 2.5 years, yielding a cross-correlation coefficient of 0.56.}
\label{figure3}
\end{figure}

\section{Data}
The data used in this study were observed by the Hinode spacecraft and belong to the Hinode synoptic programme which is executed on a regular basis. Hinode is a Japanese space mission with European and US contributions launched in late 2006 \citep[for  details about the mission see][]{2007SoPh..243....3K}. In the frame of the synoptic Hinode operation programme (HOP 79), filtergrams of the Sun are taken at the disc centre with all six available broadband filters of the 0.5 m aperture solar optical telescope \citep[SOT;][]{2008SoPh..249..167T}. 
For our purposes, the G-band filtergrams acquired in the frame of the synoptic programme are the most promising data source as they facilitate the detection of MBPs. To analyse the long-term evolution of  MBP activity, we downloaded the complete synoptic image data set from the Hinode data centre \citep[DARTS\footnote{accessible via: https://darts.isas.jaxa.jp/solar/hinode/};][]{2007SoPh..243...87M} for the period from 1 November 2006
until end of August 2014 . The complete data set consists of 4 162 single G-band exposures.

At the beginning of the Hinode operations, these observations were performed with the full pixel sampling of 0.054 arcsec/pixel. Thus an image of 4096 $\times$ 2048 pixels results in roughly 220 $\times$ 110 arcsec$^2$, which corresponds to the maximum attainable Hinode broadband filtergram field of view (FOV). Because of  the failure of the main downlink antenna, towards the end of  2007, the mission operators decided to perform an ongoing onboard binning of the synoptic images  (with the last, non-binned, synoptic image  taken on 17 February 2008). Thus the pixel sampling was reduced to 0.108 arcsec/pixel or a factor of two while keeping the FOV unchanged. A typical synoptic image taken in the G-band is displayed in Fig. \ref{figure1}. In the left-hand side of the FOV we can see some remnants of a former active region. All the primary data reductions such as dark current, flat field, and read out errors have been applied to the data via the corresponding SolarSoftWare (SSW) routines provided by the mission team, leading to so-called Level 1 data. In addition, the images observed prior to the antenna failure have been rebinned before analysis to ensure similar image characteristics for the full period under consideration.

Monthly sunspot numbers were obtained from the homepage of the Sunspot Index and Long-term Solar Observations centre (SILSO; http://sidc.oma.be/silso/).

\begin{figure}
\begin{center}
\includegraphics[width=0.95\columnwidth]{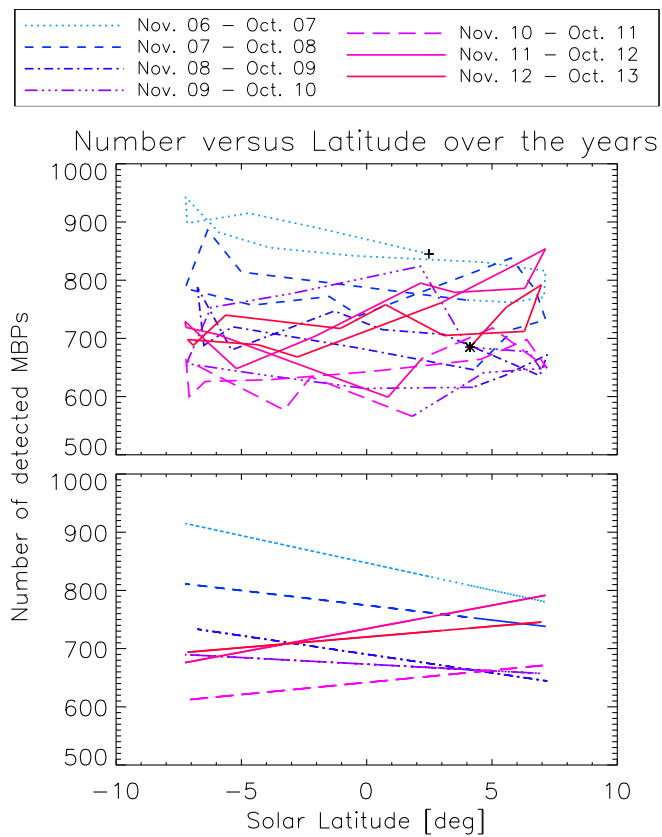}
\end{center}
\caption{Monthly median number of MBPs detected over a FOV covering approximately $\pm5^{\circ}$ in latitude versus solar latitude, plotted separately for yearly periods from November 2006 to October 2013. Top panel: The starting point is marked by a cross while the last data point is depicted as a star. Each yearly period is shown in a different color and line style (according to the legend). Bottom panel: Same as in the top panel but each yearly period is approximated by a linear trend line encoded in the same color and line style as in the top panel.}
\label{figure4}
\end{figure}

\section{Analysis}
The first step of the analysis consists of identifying the MBPs in all images. For this purpose we used two fully automated routines for image segmentation and MBP identification, as described in \citet{2009A&A...498..289U,2010A&A...511A..39U}, which always used the same set of parameters. The raw outcome of this procedure is plotted in Fig. \ref{figure2} and shows the number of identified MBPs versus  time\footnote{We  note that the absolute number of MBPs is not to be taken too seriously as it depends on the identification parameters as well as the data set (spatial resolution, spatial sampling, etc.). It is preferable to compare the relative changes over time. For a detailed study on the number density per unit surface we refer to \cite{2010ApJ...715L..26S}, with which our values could be recalibrated.}. Several features are obvious. First of all, if we discard outlier data points (MBP numbers larger than about 1100), we observe two maxima in 2006 and in 2014, and a minimum in 2011. 
In addition, we  see a cloud of scattered data points beneath this trend line. This scatter is due to erroneous images, e.g. a part of the image is missing due to downlink errors, or the image was out of focus when recorded \citep[the presence of out-of-focus images in the Hinode HOP79 was also reported by][]{2010CEAB...34...89M,2011SoPh..274...87M}. Moreover, we can identify scattered data points that exceed the level of 1100, which are often arranged in the shape of a vertical line. 
These features are created by the passage of an active region through the centre of the Sun. The strong increase can be explained by the fact that, compared to the quiet Sun,  the magnetic network is generally more enhanced within the vicinity of an active region  and visible as plage regions, which  thus gives rise to a higher number of MBPs \citep[for a more detailed discussion about the different magnetic features and pattern visible in the surrounding of sunspots, see][]{1987ARA&A..25...83Z}.

Thus we need to apply a careful image selection prior to an in-depth analysis if we aim to deduce changes in the quiet solar magnetic network as seen by the number of detected MBPs in non-active regions on the Sun. First of all we aimed to exclude faulty images\footnote{Those images for which an error occurred during the downlink from the satellite leading to  data loss and black stripes within the image.}, de-focused images, and images containing sunspots. To do this we applied a selection criterion based on the image contrast. The image contrast, $C_I$, can be defined as standard deviation of the intensity of all pixels in the image, $\sigma_I$, normalised by the mean image intensity, $\left\langle I \right\rangle $:
\begin{equation}
C_I=\frac{\sigma_I}{\left\langle I \right\rangle} \cdot 100\%,
\label{equation1}
\end{equation}
where $C_I$ is then given as the image contrast as a  percentage of the mean image intensity. 

In the next step we defined a contrast threshold and only those images within the boundary of the threshold were selected for further analysis. For this study, we  chose the values of 10.8\% as a lower and 11.8\% as an upper limit. These chosen boundaries ensure, on the one hand, that faulty images are excluded while, on the other hand,  are still wide enough to ensure the selection of enough images to obtain meaningful statistics. The possible influence of the image contrast on the detection of MBPs is discussed in detail in Sect. \ref{instr}.

\section{Results}
\subsection{Number of MBPs at the solar disc centre}
Figure \ref{figure3} shows the monthly median number of detected MBPs per image versus time as well as the relative sunspot number. We realise that a) the sunspot number does not correlate with the number of MBPs (correlation coefficient of -0.03), and that b) it might be worthwhile to shift the sunspot number with respect to the number of detected MBPs. Thus we computed the cross-correlation function between both quantities with the outcome that a shift of 2.5 years leads to a maximum correlation of 0.56 (see lower panel of Fig. \ref{figure3}). 
Furthermore, we realise that even in the deep solar minimum of Cycle 24 (during 2009 to 2011) there is still sufficient magnetic field present to give rise to about 570 detected MBPs compared to around 940 at the end of 2006, which corresponds roughly to a factor of 0.6.  

\subsection{Number of MBPs versus solar latitude close to the equator}
One has to be careful when specifying the position on the solar disc where the MBPs are detected. While it is true that during the synoptic programme the Hinode instrument is pointing to the solar disc centre, this disc centre does not always correspond to the solar equator region. This is due to the inclination of the solar rotation axis compared to the ecliptic plane given by the $B_0$ value -- the heliographic latitude of the disc centre, which changes periodically during the year while the Earth revolves around the Sun. 

To shed more light on the influence of the varying $B_0$ parameter on the detected number of MBPs, in Fig. \ref{figure4} we depict  the monthly median number of MBPs detected within a FOV that extends roughly $\pm 5^{\circ}$ in latitude, centred around the specified heliographic latitude on the $x$-axis. Each colour within the plotted curve represents a different year, starting with the  black cross in November 2006 and continuing until the black star (October 2013). We realise that the first observations were obtained north of the equator (about $3^{\circ}$), then the observations are taken more southwards, reaching roughly $-7^{\circ}$ about half a year later, before the trend in inclination changes and the instrument takes observations further northwards again. 

\begin{figure}
\begin{center}
\includegraphics[width=0.95\columnwidth]{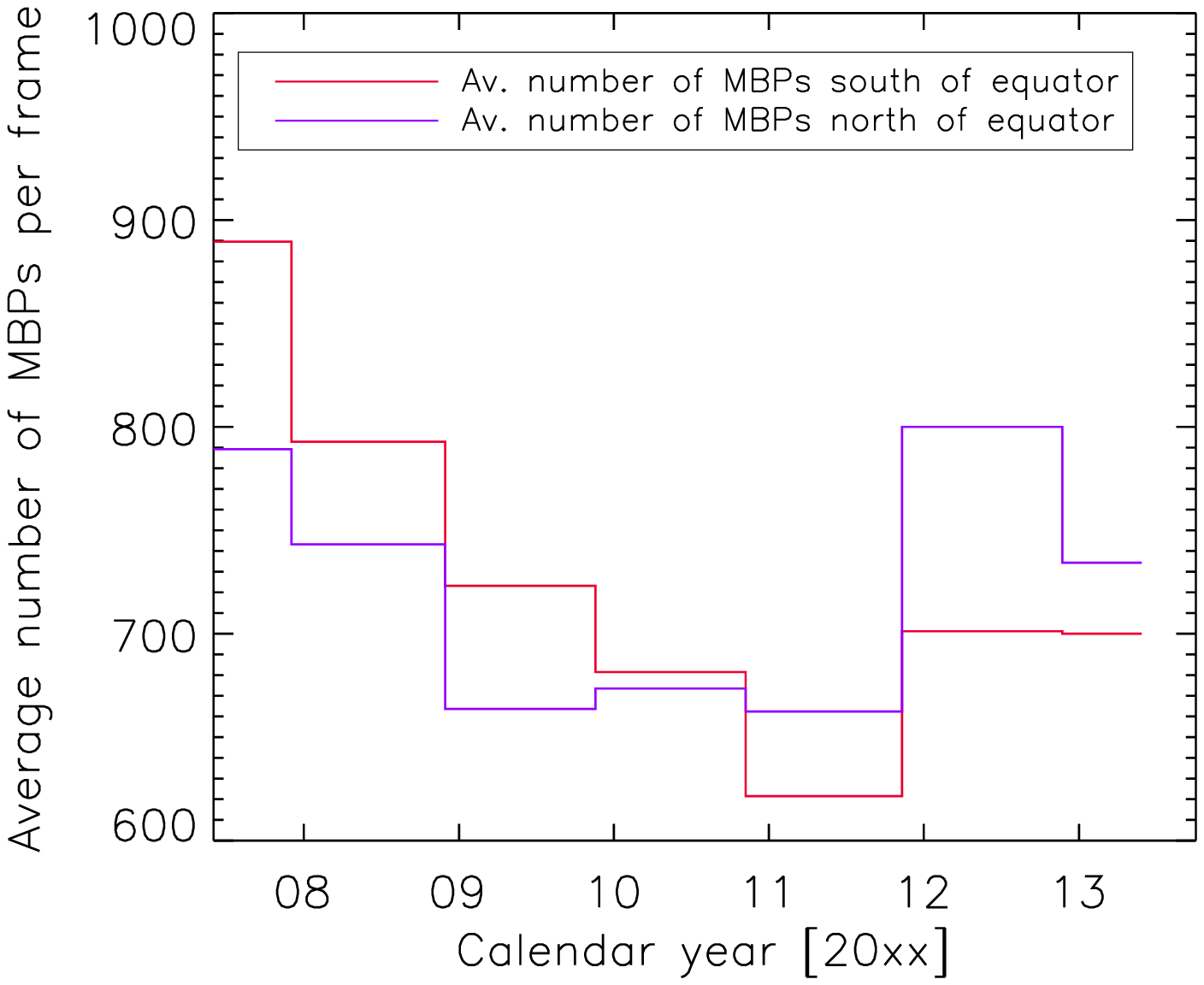}
\includegraphics[width=0.95\columnwidth]{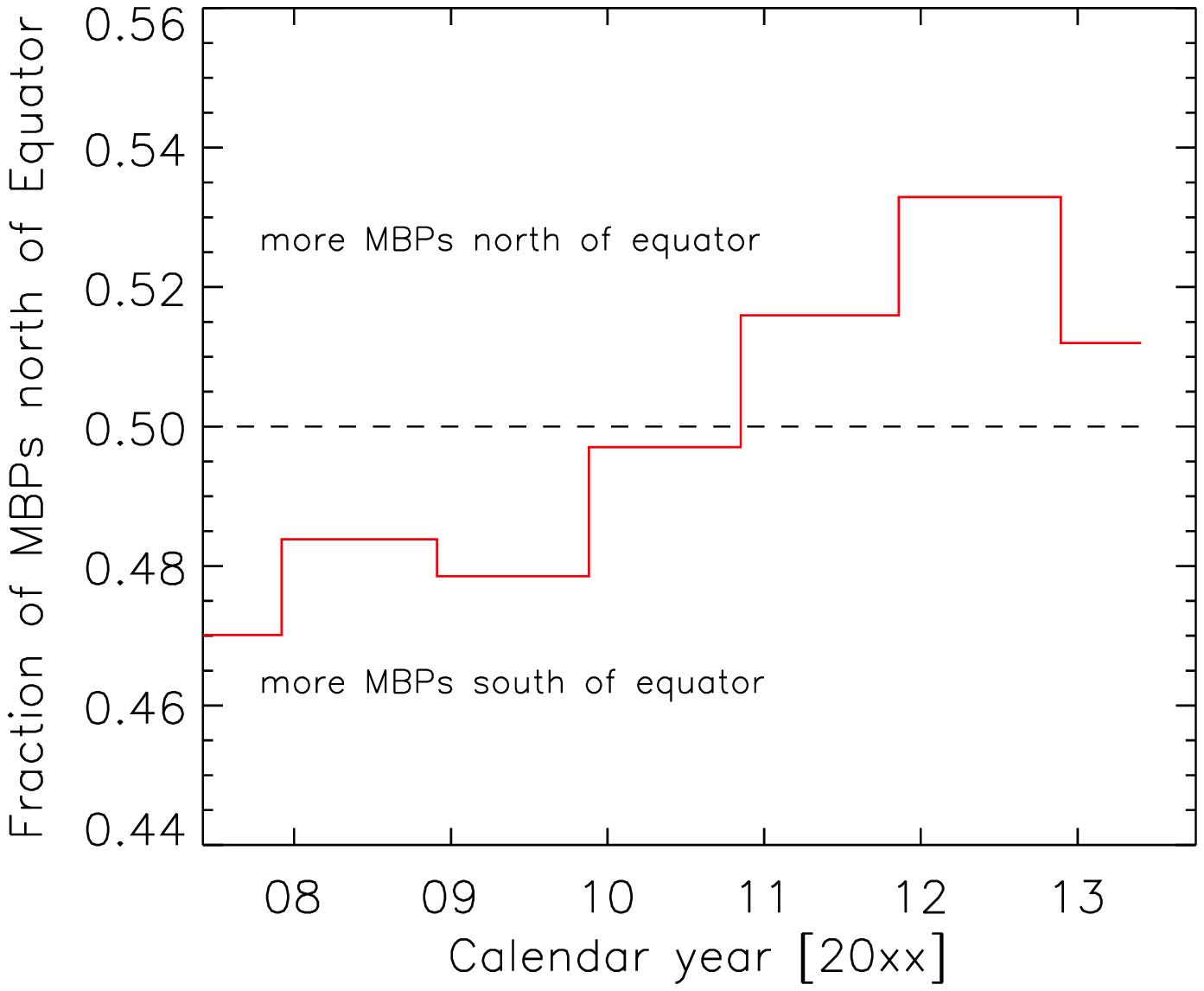}
\end{center}
\caption{Top: North/south distribution of MBPs. For each yearly period the average number of MBPs per image recorded north (blue ) and south (red ) of the solar equator is shown. Bottom: Ratio of MBPs detected north of the equator compared to the total detected number.}
\label{figure5}
\end{figure}

In the beginning, we see that the southernmost observation corresponds to a maximum in the detected number of MBPs, while the northernmost position corresponds to the yearly minimum. The following year (indicated in darker blue colour and as a dashed line) shows a similar behaviour but in general shifted to lower values. The number of detected MBPs rises again in  2011 but  now there are a higher number of detected MBPs to the north of the equator.  Additionally we are able to identify peaks in the number of identified MBPs at the covered maximum latitudes of $\pm 7^{\circ}$. In the lower panel of Fig. \ref{figure4}, we display the same plot but with linear fits for each yearly period. It shows that there are more MBPs in the southern hemisphere before the solar minimum, less MBPs after, and nearly the same amount on each side of the equator during the minimum. 

\subsection{Hemispheric MBP occurrence asymmetry}
In the following we investigate the asymmetry of the occurrence of MBPs north and south of the equator in more detail. To do  this, we average the number of detected MBPs for images obtained north and south of the equator within a year-long period. The outcome of this operation is shown in the top panel of Fig.~\ref{figure5}. The red solid line shows the average number of detected MBPs in  images obtained south of the equator, while the purple line shows the same quantity for  images taken north of the equator. Again the general picture of decreasing magnetic activity can be seen by a decreasing number of MBPs, with the minimum reached in the period between November 2010 and October 2011. This is at least true for the MBPs south of the equator. North of the equator the picture is not as clear because there is an extended minimum period that had already started  in November 2008 and which lasted until October 2011. This plot also illustrates  the reversal of the asymmetry with the rising activity of the new solar cycle (2012 onwards) when the northern hemisphere started to be predominant in  MBP activity.

This trend is even clearer in the bottom panel of Fig. \ref{figure5}, where we plot the ratio of MBPs detected north of the equator compared to the combined average number of MBPs detected north and south of the equator. In this plot the decreasing asymmetry in the south can  clearly be seen as being in line with a decreasing activity of solar cycle 24. During 2011 the asymmetry changes and the northern side is now more active with respect to the number of detected MBPs.

\section{Instrumental effects\label{instr}}
\begin{figure*}
\begin{center}
\includegraphics[width=0.9\textwidth]{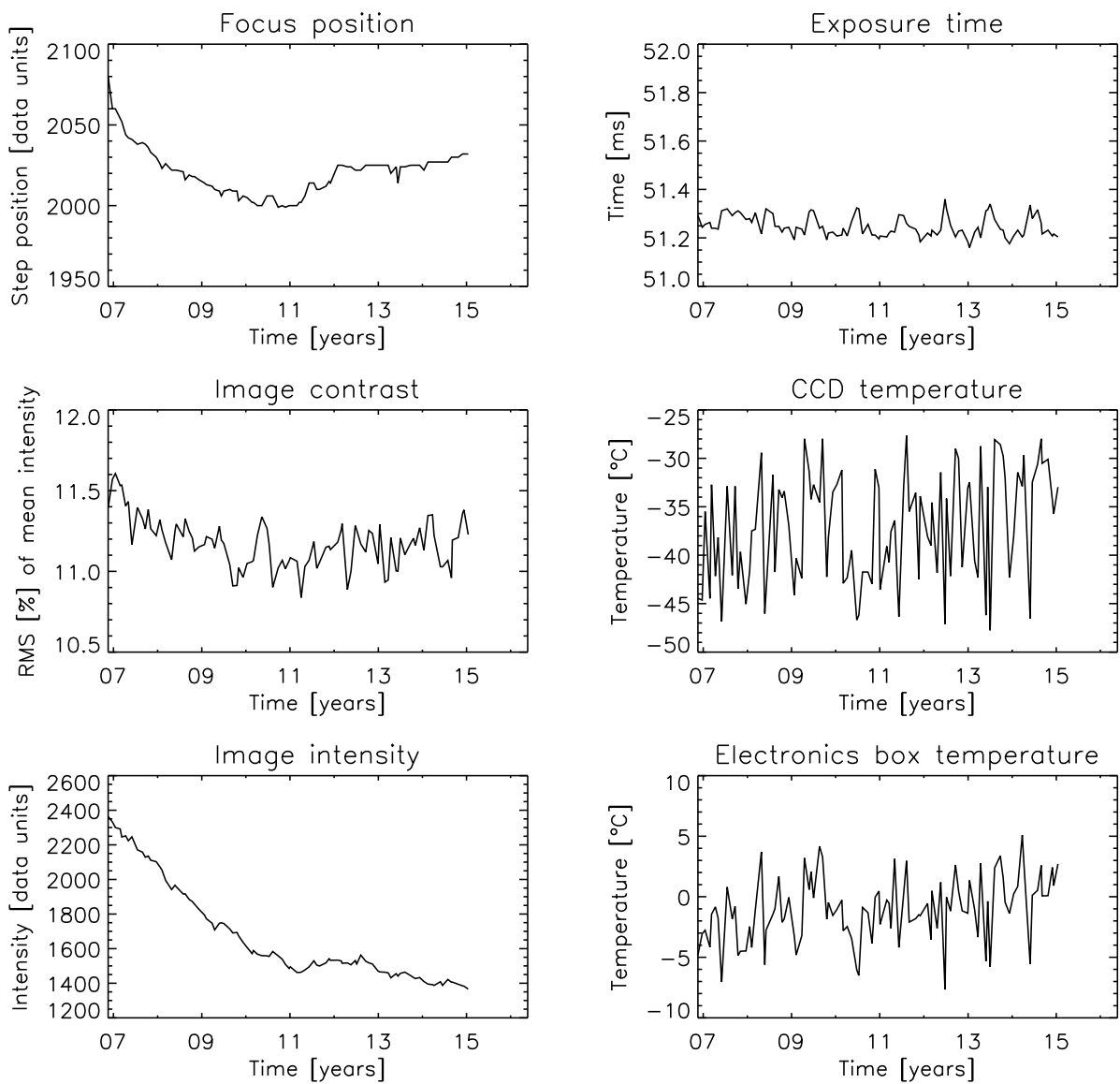}
\end{center}
\caption{Characteristic quantities describing the long-term instrument stability of the Hinode/SOT G-band filter. Left column (from top to bottom): the focus position used to record the image, the image contrast achieved, and the mean image intensity; right column: the exposure time, the CCD temperature, and the electronics box temperature during the time when the image was recorded.}
\label{figure7}
\end{figure*}
\begin{figure*}

\begin{center}
\includegraphics[width=0.78\textwidth]{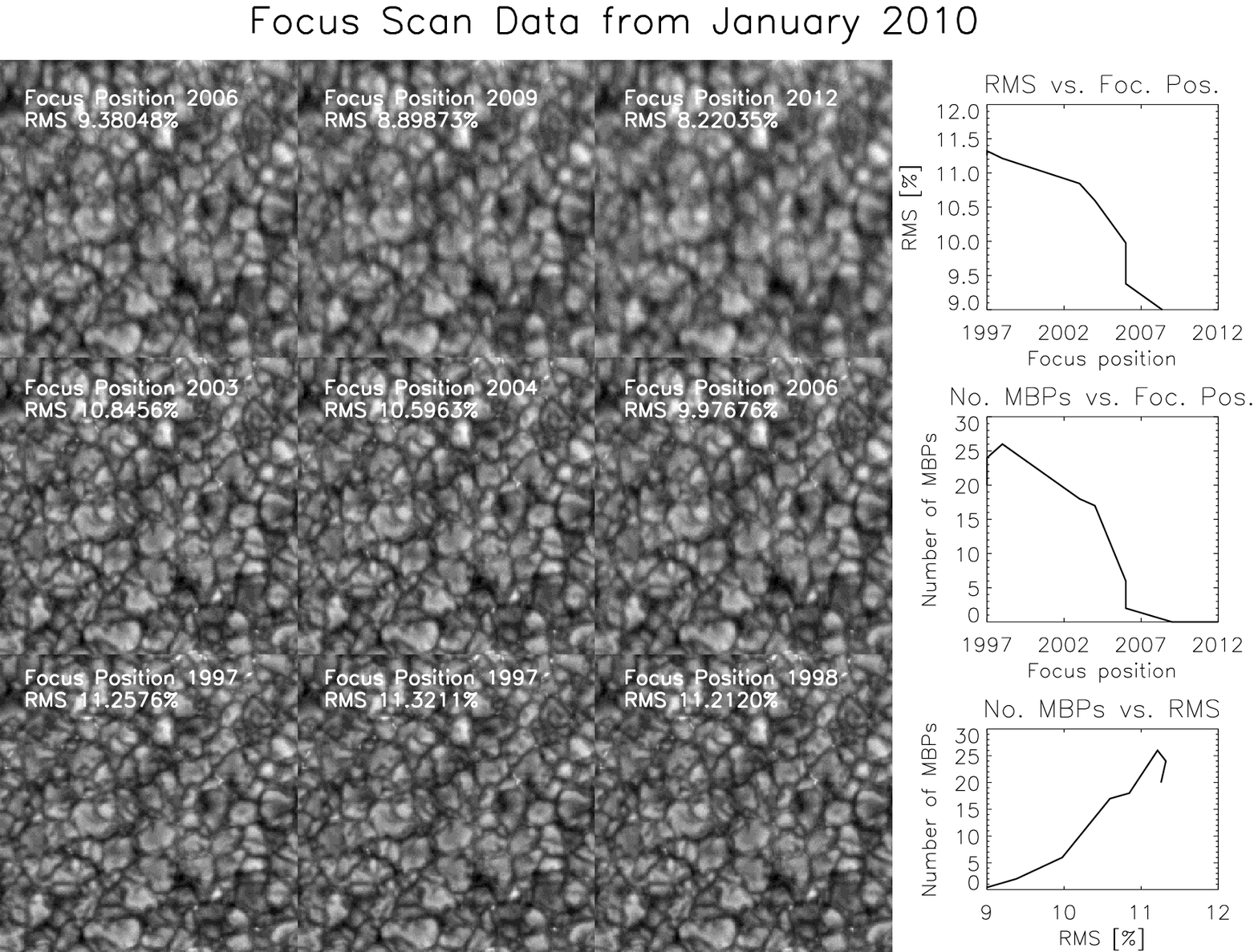}\\
\includegraphics[width=0.78\textwidth]{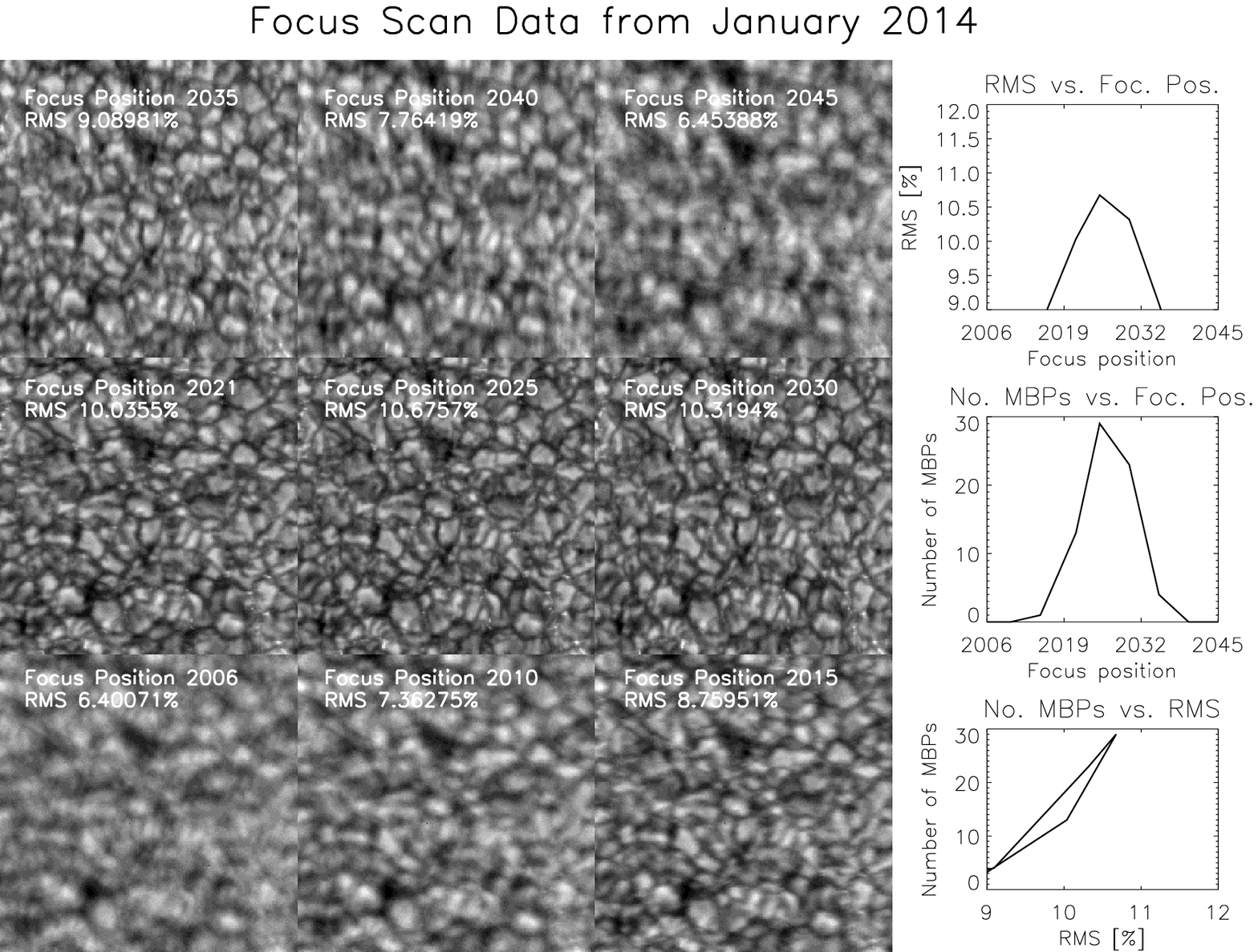}
\end{center}
\caption{Two examples of Hinode/SOT G-band focus scan data. Upper half: focus scan from January 2010. The nine-image mosaic displays the exposures recorded during the scan at different focus positions. The three panels to the right illustrate the dependence of the image contrast on the focus position, the number of detected MBPs versus the focus position, and the number of MBPs versus the image contrast. The lower half shows the same for the focus scan taken in January 2014.}
\label{figure8}
\end{figure*}

When analysing long-term trends of physical objects it is  important to have a clear understanding of the long-term behaviour of the instrument with which the data were obtained. Thus we take a closer look at the characteristic parameters of the Hinode/SOT instrument and the obtained G-band images to understand and verify if the trends obtained are of a true physical nature or an artificial product of changing instrumental characteristics.

Figure \ref{figure7} gives an overview of the long-term evolution of characteristic image parameters. The left column shows (from top to bottom) the focus position of the instrument, the image contrast as defined in Eq. \ref{equation1}, and the mean image intensity; the right column shows the exposure time, the charge-coupled device (CCD) temperature, and the electronics box temperature. All quantities are plotted versus time, thus characterising the long-term evolution of the instrument. It becomes clear that the exposure time, CCD temperature, and electronics box temperature are (within some fluctuations) stable quantities without a visible long-term trend.

However, this does not hold for the quantities plotted in the left column: the focus position, mean image intensity, and the contrast of the image, which a) reveal long-term trends and b) are correlated with each other. Most likely two effects are taking place: an ageing of the telescope structure leading to a shift in the optimum focal point\footnote{This was actually acknowledged by the satellite operators and is described in \citet{2008ASPC..397....5I}.}, and a UV blinding of the filters, which decreases the filter throughput. While the mean image intensity is of no great concern for the current study, the situation is quite different for the image contrast. This quantity is  important as we can assume (and will show later) that the detection algorithm is sensitive to the contrast value that is actually achieved, and thus a varying contrast can cause artificial variations in the number of detected MBPs. 

To verify that the shift in focus position is due to the efforts made to compensate the structural changes that occur because of ageing effects, we obtained so-called focus scan data sets. To estimate the optimum focus position, the satellite operators of the Hinode mission perform focus scans at regular intervals. The optimum position can change because of  the instrument's age, but also because of changing heat loads on the structure when the pointing of the satellite within the solar disc is changed significantly, e.g. changing the pointing from the disc centre to the limb. Two examples of such focus scans are presented in Fig. \ref{figure8}. The upper half of the figure shows the focus scan example taken in January 2010. The nine sub-images display the exposures taken during the scan at different focal positions. The three panels on the right illustrate the dependence of the RMS contrast versus the focal position, the number of detected MBPs versus the focal position, and the number of MBPs versus the achieved image contrast. The lower half of the figure shows the same plots for the focus scan from January 2014.

From Fig. \ref{figure8} we can deduce two facts: first of all the used algorithm is sensitive to image contrast variations, i.e. a change in image contrast of about 1\%, for example a decrease from 11\% to 10\%, would result in a reduction of a factor of $\sim$3 in detected MBPs. Secondly, the long-term trend of changing focus positions cannot be directly responsible for the image contrast decrease since a change of only ten steps in focus position would bring the instrument already completely out of focus. Thus the change of roughly 100 focus positions in Fig. \ref{figure7} must truly correspond to an ageing effect of the instrument structure. Furthermore, since the operators regularly re-focus the instrument based on the analysis of the focus scan data sets, the largest fraction of the variation of the image contrast of the selected images (except for the first year when the instrument ageing effects are strongest) should be due to physical changes occurring on the surface of the Sun and not as a result of instrument effects.

\begin{figure}
\begin{center}
\includegraphics[width=0.98\columnwidth]{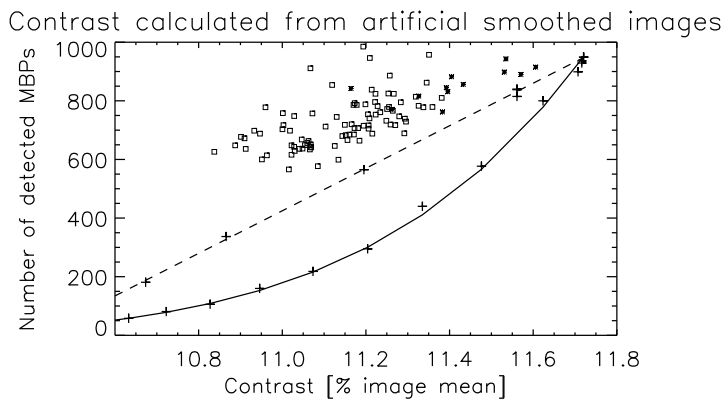}
\end{center}
\caption{The relationship between the number of identified MBPs and the image contrast obtained via artificially defocusing an image by convolving the image with a Gaussian featuring a variable width. The solid line depicts the relationship for an image taken early during the mission. Crosses demark the image contrast and the number of identified MBPs for the defocused images (always based on the same original image). The dashed line illustrates the same behaviour for a second test case using a Voigt profile for the artificial defocusing. For comparison, all analysed images (those images featuring the monthly median MBP number) from the original data set are plotted, with stars depicting the first year (strongest instrument ageing) and squares for the rest of the analysed time period.}
\label{figure9}
\end{figure}

To investigate in more detail the influence of the contrast on the detection probability of the used algorithm, we identify MBPs in artificially defocused images. We took one of the sharpest images available within the whole data set. This image was observed at the beginning of the mission before any substantial ageing effects had occurred. We analysed the image and identified roughly 940 MBPs in this particular image. In the next step, we applied a two-dimensional Gaussian filter with variable width to the image to reduce the sharpness and thus the contrast of the image. Then we reran the MBP identification algorithm and ended up, after several  such iterations, using variable Gaussian widths, with a curve depicting the performance of the identification algorithm on the achieved image contrast (Fig. \ref{figure9}, crosses). The solid black line gives a fit to the observed dependence for one particular image. The different artificially created realisations of the defocused image and the corresponding number of detected MBPs are represented by crosses. We performed the same test a second time but making a more realistic assumption for the varying telescope point spread function, where we convolved the image by a varying Voigt profile. The starting coefficients for the Voigt profile were chosen as $\sigma=0\arcsec.008$ and $\gamma=0\arcsec.004$ and varied consecutively under the constraint of keeping the ratio between both fixed. This particular function and the coefficients were derived by \citet{2008A&A...487..399W} as best matching function for the Hinode/SOT stray-light contribution in the blue continuum. As there are no parameters available for the G-band, this blue continuum profile is the closest available and most realistic profile to work with. The Voigt function used is given as Eq. 2 in \citet{2008A&A...487..399W}. The resulting curve (Fig. \ref{figure9}, dashed line) shows the same basic behaviour of a decrease in detection probability with decreasing contrast. However, we note that, in the case of the Voigt profile, the decrease is not as steep as in the case of the Gaussian function.

By way of comparison, in Fig. \ref{figure9} we overplotted  the actual contrast and number of identified MBPs of the analysed images corresponding onto the monthly median MBP numbers (stars for the first year and squares for the other years). Clearly this scatter cloud features a lower dependence of the number of MBPs versus the image contrast compared to the theoretical expectations from our modeling efforts. Hence we conclude that most images are indeed recorded at the best available focus position or at least very close to an optimum focal position. If this was not true, the scatter cloud would be characterised by a much steeper slope. Most likely the scatter is created by a small amount of residual defocused or not perfectly focused images plus the natural variation in image contrast owing to changing physical conditions on the solar surface itself. The other fact we see is that the sharpest images were obtained in the first mission year (stars). A possible explanation for this might be a decreasing CCD performance by, for example, increasing dark currents that cause a decrease in observed image RMS. This would ultimately also decrease the quantity we used to specify the sharpness of the images, while the true image sharpness would indeed show the same level.

\begin{figure}
\begin{center}
\includegraphics[height=0.89\textheight]{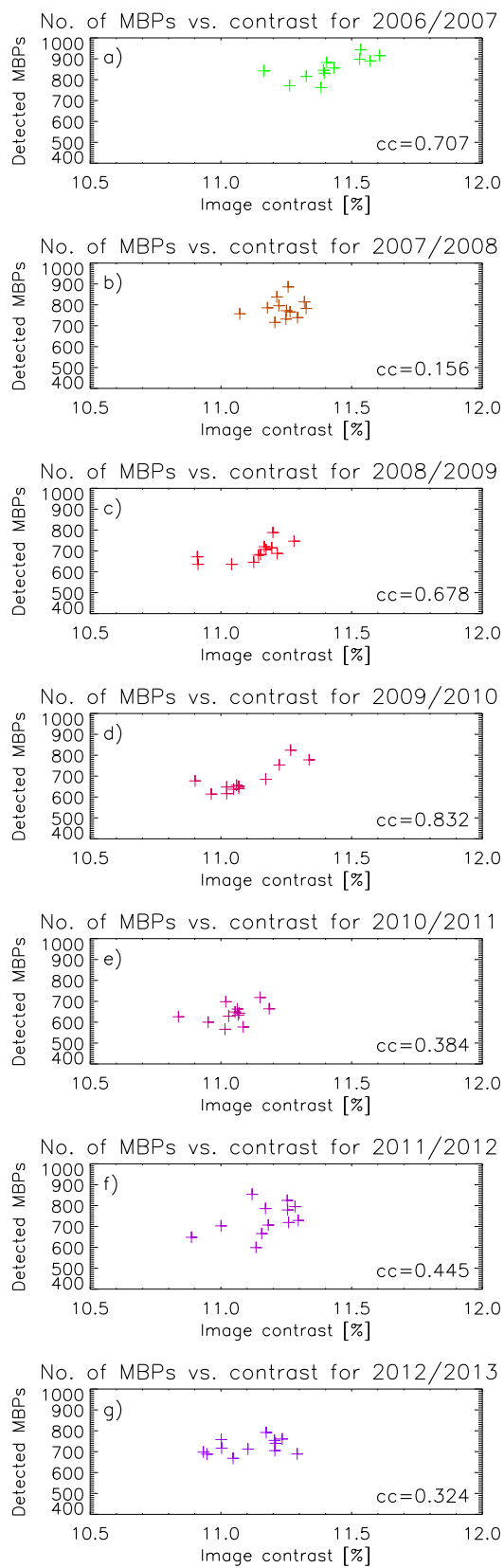}
\end{center}
\caption{Scatter plots of the monthly median number of detected MBPs per image versus image contrast, separately plotted for successive periods of 1 year. Each period is from November to October of the following year. The correlation coefficient between the number of detected MBPs and the image contrast is annotated in the lower right corner.}
\label{figure10}
\end{figure}

To achieve a better understanding of whether there are long-term variations in the data shown in Fig.~\ref{figure9}, we create separate scatter plots for yearly periods. Figure~\ref{figure10} shows (from top to bottom): the number of identified MBPs per image versus the image contrast for yearly periods starting with November 2006 -  October 2007 to  November 2012 - October 2013. There is neither a unique yearly pattern visible nor a clear correlation between image contrast and detected number of MBPs. Panel (b) and (e) display a random pattern with no clear visible correlation. The last pattern (g) would even suggest an independence of the detected number of MBPs versus image contrast as the scatter is practically arranged  along a horizontal line. The correlation coefficients are stated in the panels.

\section{Discussion and Conclusions}
The main results of this study can be summarised as follows:
\begin{itemize}
        \item The small-scale quiet Sun magnetic flux, as seen by MBPs as proxies, follows the solar cycle defined by the sunspot number.
        \item This established quiet Sun magnetic flux cycle  lags behind the sunspot cycle by a period of roughly 2.5 years.
        \item There is an asymmetry and phase shift between the number of MBPs detected north and south of the equator, and thus probably also between the magnetic network of the two hemispheres.
        \item The asymmetry switches with the start of cycle  24 and thus  the magnetic network also shows a more active or more prominent hemisphere than known for sunspots \citep[cf.][]{2006A&A...447..735T}.
        \item The Hinode/SOT shows ageing effects with regard to the optimum focus position as well as filter throughput.
        \item However, because of regular re-focusing operations, the effects on the image contrast can be successfully minimised.
\end{itemize}

These results can be interpreted in the following way: The temporal shift between the sunspot number and the number of MBPs established during the minimum periods of both quantities can be explained either by a possible transport of the magnetic fields from the sunspot belts to the equator region causing the observed temporal lag or, secondly, by a larger decaying time constant for the magnetic network compared to sunspots. However, as the meridional large circulation pattern generally transports magnetic flux from equator regions to the poles, it is most likely some different mechanism which must be acting on the MBPs in the case of the transport scenario. We can think of two possibilities: a) on small-scales, such as the few degrees distance of the active sunspot belts to the equator, the plasma flow resulting from the supergranular pattern with typical speeds of $\sim$ 300 m s$^{-1}$ can overcome the meridional flow pattern which has speeds in the order of 10--20 m s$^{-1}$ and which are strongest at latitudes of $\sim30^{\circ} $ \citep{2010ApJ...725.1082H,2010Sci...327.1350H}, or that b) close to the solar equator, a weak reverse flow exists \citep[similar to the two-part vertical circulation pattern described in][]{2013ApJ...774L..29Z}, which  transports magnetic flux from higher latitudes towards the poles, while close to the equator the magnetic flux is moved towards the equator. A different interpretation could be that there is no real time shift but that, when the feed in of new magnetic flux into the magnetic network by decaying active regions stops (during the minimum of the sunspot cycle), the magnetic network needs more time to adjust (decrease its activity), compared to sunspots decaying within days to weeks \citep[e.g.][]{1974MNRAS.169...35M}. Such an interpretation is supported by the work of \citet[][]{2014ApJ...796...19T} who find a temporal lag between the sunspot cycle and the filling factor of the magnetic network of roughly 2.5 years. In their simulation they use observed sunspot data as input into a Monte Carlo simulation of the evolution of the full disk magnetic network \citep[for more details about the simulation see also][]{2012ApJ...757..187T}. However, as their stated magnetic filling factor evolution is for the full solar disk,  no one-to-one correlation exists for the observations made in this paper. On the other hand, when the cycle starts again, the magnetic network might get enhanced simultaneously with the rising phase of the new sunspot cycle \citep[see also][Fig. 12]{2014ApJ...796...19T}. A decision, as to which possible interpretation of the observed time lag (transport or relaxation time lag) is more suitable, may be found when the period of the next solar cycle maximum is analysed. If both maxima (number of detected MBPs and sunspots) occur simultaneously then the observed behaviour is due to a larger decay time, which is constant for the network compared to sunspots. If there is a similar time lag at the maximum periods,  i.e. during the minimum phase, then it is most likely related to flux transport mechanisms. 

The number of about 570 MBPs during the extended solar cycle minimum might either be related to a surface dynamo creating magnetic fields independently of the solar cycle or be the result of weak surface remnants of the magnetic fields that were created by the global solar dynamo action. \citet[][]{2014ApJ...796...19T} found, for example, that when an injection of new flux into the magnetic network is stopped, the filling factor of the magnetic network only decreases exponentially with an e-folding time of about 1.9 years. Thus, they concluded that the magnetic network was unable to fully relax to its ground state even during the extended period of the last solar cycle around 2009. On the other hand \citet[][]{2014ApJ...797...49G} found that the whole network magnetic field can be sustained, at least during the last extended minimum solar cycle phase, via transported and accumulated weak internetwork fields. Thus the found minimum number and/or the ratio with the maximum number of detected MBPs during the maximum phase can help us, in the future, to estimate the strength of the acting surface dynamo in relation to the global magnetic field dynamo and/or the decaying time constant for the global magnetic fields. The peaks in the number of detected MBPs close to $\pm7^{\circ}$ of solar latitude are an indication that at least a considerable proportion of the magnetic field, visible as MBPs, consists of remnants of former active regions \citep[see, e.g.][for the transport of magnetic flux away from sunspots]{1973SoPh...28...61H}. This can be explained by the fact that the active sunspot belts can move on average as close to the equator as $\pm 8^{\circ}$ \citep[see][]{2003ApJ...589..665H} and thus enhance the magnetic network at these latitudes while decaying. We state ``former active regions'' as we have explicitly removed sunspot data from the data set. 

\citet{2011ApJ...731...37J}, and \citet{2012ApJ...745...39J} report about the temporal evolution of the magnetic network during solar cycle 23 using Soho/MDI observations. In their studies they divide the magnetic network elements into three classes according to their magnetic flux strength, and found that the class showing the strongest magnetic fluxes is correlated to the solar cycle, the class with weaker fluxes is anti-correlated \citep[see, for example, ][]{2003ApJ...584.1107H}, whereas the class containing the weakest flux elements reveals no correlation at all. Because of the lower spatial resolution of Soho/MDI, the weakest class of reported magnetic flux elements of this study might in reality belong to internetwork fields. Recent results for the internetwork magnetic fields \citep[see,][]{2013A&A...555A..33B,2014PASJ...66S...4L} support the idea of a working small-scale surface dynamo being responsible for these internetwork fields, which are thus also not related and correlated to the global dynamo. As we found a correlation of the MBP number activity with the solar cycle (although temporally lagged), the observed MBPs are likely created within elements belonging to the class of the stronger network magnetic flux elements. In the future, the implementation of an algorithm,  which enables a differentiation between magnetic network elements and internetwork elements within our detection algorithm, might further help to discriminate between the global and a possible acting local solar dynamo creating internetwork fields that are not correlated or even anti-correlated \citep[see][]{2015arXiv150506632K} to the solar sunspot cycle. 
To make sure that the results and interpretation are valid, we checked carefully the implications of possible ageing effects of the instrument and thus changing observational conditions on the results. From the analysis of these long-term image characteristics we conclude that the observed temporal evolution,  
as well as the asymmetry between regions north and south of the equator,
are real and caused by physical effects. There might be some instrumental effects, like a long-term degradation, which cannot be excluded and which might in turn change the slope of the evolutionary curve (strengthening the decreasing trend up to 2011 and then weakening the increasing trend of the new solar cycle). Thus the absolute numbers as well as the absolute changes might be liable to errors, but the qualitative trend, as well as the asymmetry, are persistent.

In a follow-up study, we plan to investigate the long-term evolution of the size and the contrast of the MBPs to investigate if the features themselves also vary over longer periods, or if these small-scale building blocks of the solar magnetism show stable and static characteristics. It will be worth revisiting this kind of investigation after a longer period when Hinode will have gathered even more data, hopefully covering the complete ongoing solar activity cycle, especially the maximum as this will enable us to distinguish between the two possible interpretations of the time-lag (transport effects vs. decay time constant) as outlined above.

\begin{acknowledgements}
The research was funded by the Austrian Science Fund (FWF): P20762, P23618 and J3176. Moreover, this work was 
 supported by COST Action ES1005 (TOSCA). We are
grateful to the Hinode team for the
opportunity to use their data.
Hinode is a Japanese mission developed and launched by ISAS/JAXA, collaborating with NAOJ as a domestic partner, NASA and STFC (UK) as international partners. Scientific operation of the Hinode mission is conducted by the Hinode science team organised at ISAS/JAXA. This team mainly consists of scientists from institutes in the partner countries. Support for the post-launch operation is provided by JAXA and NAOJ (Japan), STFC (U.K.), NASA (U.S.A.), ESA, and NSC (Norway).
D. U. and A. H. are grateful to the {\"O}AD ({\"O}sterreichischer Austauschdienst) for financing a
scientific stay at the Pic du Midi Observatory. R.M. is grateful to the Minist\`{e}re des Affaires Etrang\`{e}res et Europ\'{e}ennes
for financing a stay at the University of Graz. Furthermore, D. U. wishes to acknowledge again the {\"O}AD for providing financial means to conduct short research stays at the Astronomical Institute of the Czech Academy of Sciences as well as at the Slovak counterpart. Finally M. B\'{a}rta is grateful to the M\v{S}MT for funding a research stay in Austria. Partial funding has also been obtained from the Spanish Ministerio de Econom\'{i}a through project ESP2013-47349-C6-1-R. Last but not least, all the authors wish to express their gratitude to the anonymous referee who helped  to improve this study.

\end{acknowledgements}
\bibliographystyle{aa}
\bibliography{Literaturverzeichnis}

\end{document}